\documentclass[aps,prb,amsmath,amssymb,amsfont,showpacs,superscriptaddress,reprint]{revtex4-1}

\usepackage{graphicx}
\usepackage[usenames,dvips]{color}
\usepackage{times}
\usepackage{natbib}
\usepackage{subfigure}
\usepackage{soul}

\begin{document}

\newcommand{\dd}[1]{\mathrm{d}#1}
\newcommand{\kb}{k_\text{B}}
\newcommand{\Td}{T_\text{D}}
\newcommand{\comment}[2]{#2}
\newcommand{\todo}[1]{{\color{red}#1}}

\newcommand{\JO}[1]{\color{blue}#1}

%\title{The empirical Monod-Beuneu relation revisited for elemental metals}
\title{The Elliott-Yafet theory of spin relaxation generalized for large spin-orbit coupling
%The effect of large spin-orbit coupling on the Elliott-Yafet theory of spin-relaxation
%\\or\\
}

\author{Annam\'aria Kiss}
\affiliation{Institute for Solid State Physics and Optics, Wigner Research Centre for Physics, Hungarian Academy of Sciences, POBox 49, H-1525 Budapest, Hungary}
\affiliation{BME-MTA Exotic Quantum Phases Research Group, Budapest University of Technology and Economics, Budapest, Hungary}

\author{L\'enard Szolnoki}
\affiliation{Department of Physics, Budapest University of Technology and Economics and
MTA-BME Lend\"{u}let Spintronics Research Group (PROSPIN), POBox 91, H-1521 Budapest, Hungary}

\author{Ferenc Simon}
\email[Corresponding author: ]{f.simon@eik.bme.hu}
\affiliation{Department of Physics, Budapest University of Technology and Economics and
MTA-BME Lend\"{u}let Spintronics Research Group (PROSPIN), POBox 91, H-1521 Budapest, Hungary}

%76.30.Pk Electron paramagnetic resonance and relaxation Conduction electrons,(BULK)	 71.70.Ej Spin-orbit coupling, Zeeman and Stark splitting, Jahn-Teller effect, 72.25.Rb	 Spin relaxation and scattering 75.76.+j	Spin transport effects (for devices exploiting spin polarized transport, see 85.75.Hh, 85.75.Mm, and 85.75.Ss)

\date{\today}

\begin{abstract}
We generalize the Elliott-Yafet (EY) theory of spin relaxation in metals with inversion symmetry for the case of large spin-orbit coupling (SOC). The EY theory treats the SOC to the lowest order but this approach breaks down for metals of heavy elements (such as e.g. caesium or gold), where the SOC energy is comparable to the relevant band-band separation energies. The generalized theory is presented for a four-band model system without band dispersion, where analytic formulae are attainable for arbitrary SOC for the relation between the momentum- and spin-relaxation rates. As an extended description, we also consider an empirical pseudopotential approximation where SOC is deduced from the band potential (apart from an empirical scaling constant) and the spin-relaxation rate can be obtained numerically. Both approaches recover the usual EY theory for weak SOC and give that the spin-relaxation rate approaches the momentum-relaxation rate in the limit of strong SOC. We argue that this limit is realized in gold by analyzing spin relaxation data. A calculation of the $g$-factor shows that the	empirical Elliott-relation, which links the $g$-factor and spin-relaxation rate, is retained even for strong SOC.
\end{abstract}

\maketitle

%\section{Introduction}
The prospect of using electron spins as information carriers, (a field known as spintronics) renewed the interest in the fundamental desciption of spin relaxation in metals and semiconductors. Spintronics devices rely on the controlled creation and readout of a non-equilibrium net spin population. Spin relaxation in turn characterizes how rapidly the non-equilibrium spin population decays, knowledge and theoretical description of spin relaxation is therefore of central importance.

The first experiment in the field dates back to 1955 (Ref. \onlinecite{FeherKip}), which reported the first electron spin resonance in metals. The first proper theoretical description of spin relaxation in metals with inversion symmetry was provided by Elliott \cite{Elliott}, which was later generalized to lower temperatures and for various relaxation mechanisms by Yafet \cite{yafet1963g}. The Elliott-Yafet (EY) theory of spin relaxation is valid for metals and semiconductors (metals in the following) with i) inversion symmetry, ii) weak spin-orbit coupling (SOC), and iii) low quasi-particle scattering rate \cite{FabianRMP,WuReview}. When the inversion symmetry is broken, the so-called D'yakonov-Perel' mechanism describes the spin relaxation \cite{DyakonovPerelSPSS1972}. The case of sizeable quasi-particle scattering was described before \cite{SimonPRL2008,DoraSimonStronglyCorrPRL2009} but strong SOC has not been considered yet.

The conventional EY theory exploits that in the presence of inversion symmetry, the spin-up and spin-down states remain degenerate as a result of time-reversal invariance (or Kramers' theorem) until the later is broken by e.g. a magnetic field. The presence of a nearby band gives rise to an admixture of the spin-up/down states in the conduction band, while the energy degenerancy is retained. As a result, the EY desciption is a four-band theory. The admixed states read:

\begin{eqnarray}
{| \widetilde{\uparrow} \rangle}_{\boldsymbol k} &=& \left[a_{\boldsymbol k}\left({\boldsymbol r}\right) | \uparrow \rangle + b_{\boldsymbol k}\left({\boldsymbol r}\right) | \downarrow \rangle\right]e^{i{\boldsymbol k}\cdot{\boldsymbol r}}, \label{eq-st1}\\
{| \widetilde{\downarrow} \rangle}_{\boldsymbol k} &=& \left[a^{\ast}_{-{\boldsymbol k}}\left({\boldsymbol r}\right) | \downarrow \rangle - b^{\ast}_{-{\boldsymbol k}}\left({\boldsymbol r}\right) | \uparrow \rangle\right]e^{i{\boldsymbol k}\cdot{\boldsymbol r}},\label{eq-st2}
\end{eqnarray}

\noindent where ${| \uparrow \rangle}$ and ${| \downarrow \rangle}$ are the pure (unperturbed) spin states and ${| \widetilde{\uparrow} \rangle}_{\boldsymbol k}$, ${| \widetilde{\downarrow} \rangle}_{\boldsymbol k}$ are the perturbed Bloch states. In the first order of the SOC, the coefficients are given by the $L$ matrix element of the SOC for the conduction and the near lying band, and the corresponding energy separation $\Delta$ as:
%$\frac{b_\mathbf{k}}{a_\mathbf{k}}\propto \frac{L}{\Delta E}$,
%ANI - ver16
%{
%$\frac{|b_\mathbf{k}|}{|a_\mathbf{k}|}\propto \frac{L}{\Delta}$.
%}
$ |b_{\boldsymbol k}| / |a_{\boldsymbol k}| \approx L/\Delta$.

Elliott showed with a first-order time dependent perturbation treatment \cite{Elliott} that the usual momentum scattering induces spin transitions for the admixed states, i.e. a spin relaxation. With $\Gamma_{\text{s}}=\hbar/\tau_{\text{s}}$ and $\Gamma=\hbar/\tau$ used for the spin- and momentum-relaxation rates ($\tau_{\text{s}}$ and $\tau$ are the corresponding relaxation times), respectively:

\begin{gather}
\Gamma_{\text{s}}=\alpha_1 \left( \frac{L}{\Delta}\right)^2 \Gamma,
\label{elliottrel1}
\end{gather}

\noindent where $\alpha_1$ is a band structure dependent constant near unity.

Elliott further showed that the magnetic energy of the admixed states is different from that of the pure spin-states, i.e. there is a shift in the electron $g$-factor:

\begin{gather}
\Delta g=g-g_0=\alpha_2 \frac{L}{\Delta},
\label{elliottrel2}
\end{gather}

\noindent where $g_0 \approx 2.0023$ is the free electron $g$-factor, $\alpha_{2}$ is another band structure dependent constant near unity. Eqs. \eqref{elliottrel1} and \eqref{elliottrel2} gives the so-called Elliott relation

\begin{equation}
\Gamma_{\text{s}}=\frac{\alpha_1}{\alpha_2^2}\Delta g^2 \Gamma,
\label{elliottrel3}
\end{equation}

\noindent which links three empirical measurables; $\Gamma_{\text{s}}$, $\Gamma$, and $\Delta g$. In practice, the spin-relaxation time is obtained for metals from conduction electron spin resonance (CESR) measurements \cite{FeherKip} as: $\Gamma_{\text{s}}=\hbar\gamma\Delta B$. Here $\Delta B$ is the homogeneous ESR line-width and $\gamma/2\pi= 28.0 \,\text{GHz/T}$ is the electron gyromagnetic ratio. The CESR resonance line position yields the $g$-factor shift.

Monod and Beuneu tested empirically \cite{MonodBeuneuPRB1979} the validity of Eq. \eqref{elliottrel1} and found that the atomic SOC induced energy splittings approximate well the appropriate matrix elements. For most elemental metals
%$\frac{L}{\Delta E}\ll\,0.1$
$L / \Delta \ll\,0.1$
holds but for Cs, Cu, Ag, and In it is about 0.1 and for other metals 0.2 (Pb), 0.3 (Hg and Sn) with a 0.9 as the largest value in Au. Clearly, the validity of a first-order perturbation treatment of the SOC for these cases is questionable. This motivates us to revisit the EY theory for the case when the SOC is not small compared to other energy scales (kinetic energy or band separations).

Here, we discuss the most general form of the SOC Hamiltonian which is applicable for the EY model. We proceed with a simplified four-band Hamiltonian and solve the problem of spin relaxation in the first order of the scattering but exactly for the SOC. We find that the conventional EY result is recovered for weak SOC. A calculation of the $g$-factor yields the Elliott-relation for arbitrary strength of the SOC. A numerical calculation is also presented for the spin relaxation in the framework of a pseudopotential approximation for the conduction electrons, where SOC is obtained directly from the potential together with an empirical scaling factor. The numerical result qualitatively returns the result of the EY model calculations. The most important prediction is that the spin scattering time approaches the momentum scattering for strong SOC. We revisit previous experimental data on Au by Monod and J\'anossy \cite{MonodJanossy1977} and show that this effect was observed already but it avoided the attention.

%{\color{red}
\section*{The Elliott theory of spin relaxation}%}

We consider the Elliott mechanism of spin relaxation\cite{Elliott} in which the conduction electron spin interacts with its motion in the electric field of the host lattice described by the periodic potential $V_{\ell}$. This lattice induced spin-orbit coupling, given by the Hamiltonian
\begin{eqnarray}
{\cal H}_{\rm SOC} &= \frac{\hbar^2}{4m^2c^2} (\nabla V_{\ell} \times {\boldsymbol k})    \cdot  {\boldsymbol \sigma},
\label{h-soelliott}
\end{eqnarray}
leads to the mixing of the originally pure spin states in the electron wave function.
Thus, the conduction electron spins can now relax through ordinary momentum scattering caused by impurity atoms, for example.
The spin-relaxation rate is estimated from the spin-flip transition probability $W_{{\boldsymbol k} \sigma \rightarrow {\boldsymbol k}^{\prime} \sigma^{\prime}}$ of an electron scattering on the impurity potential $V$, and similarly the momentum-relaxation rate from the non spin-flip transition probability $W_{{\boldsymbol k}\sigma \rightarrow {\boldsymbol k}^{\prime} \sigma}$.
In this picture we neglect the SOC caused by the electric field of the impurity atoms, and also assume that they give rise to much larger momentum scattering than the host ions.

To formulate the above described Elliott mechanism, we start from the wave functions of electrons in the periodic potential $V_{\ell}$, which are Bloch-type as
\begin{eqnarray}
|\psi_{{\boldsymbol k},\sigma, n}  \rangle = | u_{{\boldsymbol k},\sigma, n} \rangle {\rm e}^{i{\boldsymbol k} \cdot{\boldsymbol r}},
\label{eq-wavef1}
\end{eqnarray}
where $n$ is the band index, and the Bloch functions $| u_{{\boldsymbol k},\sigma, n} \rangle$ are lattice-periodic. Each band is at least twofold degenerate due to the presence of time-reversal symmetry.

The spin-flip transition probability $W_{{\boldsymbol k} \sigma \rightarrow {\boldsymbol k}^{\prime} \sigma^{\prime}}$ is determined by the matrix element $ \langle \psi_{{\boldsymbol k},\sigma, n} | V | \psi_{{\boldsymbol k}^{\prime},\sigma^{\prime}, n} \rangle $ within a first-order perturbation theory, and in parallel, the non spin-flip transition element $W_{{\boldsymbol k} \sigma \rightarrow {\boldsymbol k}^{\prime} \sigma}$ is related to the matrix element $ \langle \psi_{{\boldsymbol k},\sigma, n} | V | \psi_{{\boldsymbol k}^{\prime},\sigma, n} \rangle$.
If we assume that the impurity scattering potential $V$ is slowly varying on the scale of the unit cell, the matrix elements can be approximated by the factorization
\begin{eqnarray}
\langle \psi_{{\boldsymbol k},\sigma, n} | V | \psi_{{\boldsymbol k}^{\prime},\sigma^{\prime}(\sigma), n} \rangle  \approx V_{{\boldsymbol k}{\boldsymbol k}^{\prime}} \langle u_{{\boldsymbol k},\sigma, n} | u_{{\boldsymbol k}^{\prime},\sigma^{\prime}(\sigma), n} \rangle
\end{eqnarray}
with $V_{{\boldsymbol k}{\boldsymbol k}^{\prime}} = \int {\rm e}^{i({\boldsymbol k}-{\boldsymbol k}^{\prime}) \cdot{\boldsymbol r}} V({\boldsymbol r}) d^3 r$, which gives the spin-flip and non spin-flip transition elements as
\begin{eqnarray}
W_{{\boldsymbol k} \sigma \rightarrow {\boldsymbol k}^{\prime} \sigma^{\prime}}^{(n)} &=&
\frac{2\pi}{\hbar} \delta(E_{{\boldsymbol k}} - E_{{\boldsymbol k}^{\prime}})
\left(V_{{\boldsymbol k}{\boldsymbol k}^{\prime}} \right)^2
|\langle u_{{\boldsymbol k},\sigma, n} | u_{{\boldsymbol k}^{\prime},\sigma^{\prime}, n} \rangle|^2
, \nonumber\\
\label{wuddef1}
\\
W_{{\boldsymbol k} \sigma \rightarrow {\boldsymbol k}^{\prime} \sigma}^{(n)} &=&
\frac{2\pi}{\hbar} \delta(E_{{\boldsymbol k}} - E_{{\boldsymbol k}^{\prime}})
\left(V_{{\boldsymbol k}{\boldsymbol k}^{\prime}} \right)^2
|\langle u_{{\boldsymbol k},\sigma, n} | u_{{\boldsymbol k}^{\prime},\sigma, n} \rangle|^2
\nonumber\\
\label{wuudef1}
\end{eqnarray}
within our approximation obtained by applying the Fermi's golden rule.
Finally, the spin- and momentum-relaxation rates are calculated as
\begin{eqnarray}
\Gamma_{\text{s}} &=& \int d^3 k d^3 k^{\prime} \frac{\delta({\boldsymbol k} - {\boldsymbol k}_{\rm F})}{4\pi k_{\rm F}^2} \frac{\delta({\boldsymbol k}^{\prime} - {\boldsymbol k}_{\rm F})}{4\pi k_{\rm F}^2}  W_{k \sigma \rightarrow k^{\prime} \sigma^{\prime}}^{(n)},\label{gammas}
\\
\Gamma &=& \int d^3 k d^3 k^{\prime}  \frac{\delta({\boldsymbol k} - {\boldsymbol k}_{\rm F})}{4\pi k_{\rm F}^2} \frac{\delta({\boldsymbol k}^{\prime} - {\boldsymbol k}_{\rm F})}{4\pi k_{\rm F}^2}  W_{k \sigma \rightarrow k^{\prime} \sigma}^{(n)}
\label{gamma}
\end{eqnarray}
after a double integration of the transition probabilities on the Fermi surface given with the wave vector ${\boldsymbol k}_{\rm F}$.

\section*{Results and discussion}

%\subsection{The Elliott-Yafet model Hamiltonian}

\subsection*{The Elliott-Yafet model for strong SOC}

We consider a four-band model which consists of two nearby bands each with a two-fold degeneracy with respect to the spin-up and spin-down states. The bands are separated by an energy gap $\Delta$, and we neglect the dispersion of the bands. We found that this simplified model allows to derive analytic expression for the spin-relaxation rate for arbitrary value of the spin-orbit interaction. The most general form of the spin-orbit interaction with inversion symmetry is discussed in Ref. \onlinecite{Boross2013}:
\begin{eqnarray}
\hat{\cal H}_{\rm SOC} =
\begin{pmatrix}
0 & 0 & L_{\uparrow \uparrow} & L_{\downarrow \uparrow}\\
0 & 0 & L_{\uparrow \downarrow} & L_{\downarrow \downarrow}\\
L_{\uparrow \uparrow}^{\ast} & L_{\downarrow \uparrow}^{\ast} & 0 & 0\\
L_{\uparrow \downarrow}^{\ast} & L_{\downarrow \downarrow}^{\ast} & 0 & 0 \\
\end{pmatrix}
\label{ham-so}
\end{eqnarray}
within the basis $\{ |1\uparrow \rangle, |1\downarrow \rangle, |2\uparrow \rangle, |2\downarrow \rangle \} $ where $1$ and $2$ label the bands, and $L_{\sigma \sigma^{\prime}}$ are SOC matrix elements.

For inversion and time-reversal symmetries, the SOC matrix elements satisfy:
$L_{\uparrow \uparrow} = -  L_{\downarrow \downarrow} \in \mathbb{R}$ and $L_{\uparrow \downarrow} = (L_{\downarrow \uparrow})^{\ast} \in \mathbb{C}$. The detailed group theoretical considerations are given in the Supplementary Material. For simplicity, we keep only spin-flip SOC matrix elements which connect the states $|1\sigma \rangle$ and $|2\sigma^{\prime} \rangle$ and neglect the ones which mix the same spin directions. We also consider a small Zeeman energy, $h$, to handle computational difficulties due to the doubly degenerate bands but we reinsert $h=0$ at the end of the calculations.

\begin{figure}[htp]
\begin{center}
\includegraphics[scale=1.3]{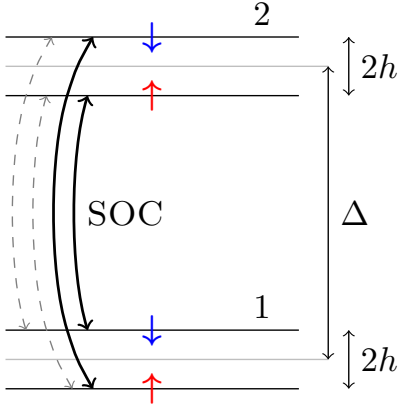}
\caption{Schematics of the four-band model with SOC and a Zeeman term, $h$.}
\label{four_band_model}
\end{center}
\end{figure}

This Hamiltonian of this simplified model is then given by:
\begin{eqnarray}
\hat{\cal H} =
\begin{pmatrix}
h & 0 & 0 & L_{\boldsymbol k}\\
0 & -h & L_{\boldsymbol k}^{\ast} & 0\\
0 & L_{\boldsymbol k} & \Delta+h  & 0\\
L_{\boldsymbol k}^{\ast} & 0 & 0 & \Delta-h \\
\end{pmatrix}
\label{ham-1}
\end{eqnarray}
which includes kinetic, SOC, and the Zeeman terms. The level structure and the relevant terms of this model are depicted in Fig.~\ref{four_band_model}.

We retain the ${\boldsymbol k}$ index for the SOC matrix elements in Eq.~(\ref{ham-1}) to indicate that it depends on both the direction and magnitude of the wave vector, even though we consider dispersionless bands. This is required as otherwise $| u_{\uparrow} \rangle$ and $| u_{\downarrow} \rangle$ are orthogonal, which would give zero spin transition rate: $W_{{\boldsymbol k} \sigma \rightarrow {\boldsymbol k}^{\prime} \sigma^{\prime}}=0$.

The original energy splitting, $\Delta$, is modified by the SOC as $\Delta({\boldsymbol k}) = \sqrt{\Delta^2 + 4 |L_{{\boldsymbol k}}|^2}$.
The Bloch functions in Eq.~(\ref{eq-wavef1}) are obtained as the eigenstates of Hamiltonian~(\ref{ham-1}):
\begin{eqnarray}
| u_{{\boldsymbol k}, \widetilde{\sigma}, n} \rangle \equiv | n \widetilde{\sigma} \rangle_{{\boldsymbol k}} = \sum_{n = 1,2} \sum_{\sigma = \uparrow, \downarrow} c_{{\boldsymbol k} n \sigma} |  n \sigma \rangle,
\end{eqnarray}
where the index $\widetilde{\sigma}$ denotes mixed spin states due to the SOC.
For example, the lowest two states are obtained as
\begin{eqnarray}
 | 1 \widetilde{\uparrow} \rangle_{{\boldsymbol k}} &=& a_{\boldsymbol k} | 1 \uparrow \rangle + b_{\boldsymbol k} | 2 \downarrow \rangle,
 \label{eq-nst1}
 \\
  | 1 \widetilde{\downarrow} \rangle_{{\boldsymbol k}} &=& a_{\boldsymbol k}^{\ast} | 1 \downarrow \rangle + b_{\boldsymbol k}^{\ast} | 2\uparrow \rangle,
 \label{eq-nst2}
\end{eqnarray}
which have the same form as the ones given in Eqs.~(\ref{eq-st1}), (\ref{eq-st2}) derived by Elliott.
The explicit expressions for the coefficients $a_{\boldsymbol k}$ and $b_{\boldsymbol k}$ in Eqs.~(\ref{eq-nst1}), (\ref{eq-nst2}) for arbitrary value of the SOC are given in the Supplementary Material. %Appendix~\ref{app:four_band}.
In the limit of small SOC, i.e. $|L_{\boldsymbol k}|/\Delta \ll {\cal O}(1)$, we recover the perturbation result of Elliott\cite{Elliott} as $a_{\boldsymbol k} \approx 1 - |L_{\boldsymbol k}|^2/(2\Delta^2) = 1 - {\cal O}(L^2)$ and $b_{\boldsymbol k} \approx |L_{\boldsymbol k}|/\Delta$.

The spin-flip and non spin-flip transition probabilities can be calculated at each band based on the formulae~(\ref{wuddef1}) and (\ref{wuudef1}) and the general result is given in the Supplementary Material. Analytic results are obtained when i) the Fermi surface is approximated with a sphere with radius $k_{\rm F}$ which is a good approximation for (${\boldsymbol k} \approx 0$), ii) the unknown overlap integrals of the orbital part of the states and the impurity potential is estimated by constants as
$  _{\boldsymbol k}\langle 1  |  1  \rangle_{\boldsymbol k^{\prime}} =\alpha_{{\boldsymbol k}{\boldsymbol k}^{\prime}} \approx \alpha$,
$  _{\boldsymbol k}\langle 2  | 2  \rangle_{\boldsymbol k^{\prime}} =\beta_{{\boldsymbol k}{\boldsymbol k}^{\prime}} \approx \beta$,
$  _{\boldsymbol k}\langle 1  | 2  \rangle_{\boldsymbol k^{\prime}} =\gamma_{{\boldsymbol k}{\boldsymbol k}^{\prime}} \approx \gamma$,
$\sqrt{2\pi/\hbar} V_{{\boldsymbol k}{\boldsymbol k}^{\prime}} \approx V$, and iii) the SOC matrix elements are approximated by their average values at the Fermi surface: $|L_{{\boldsymbol k}({\boldsymbol k}^{\prime})}| \approx |L(k_{\rm F})| \equiv L$.
The latter two approximations are justified as these quantities are band structure dependent constants after the integration over the Fermi surface such as $\alpha_{1}$ is introduced in Eq.~(\ref{elliottrel1}).

With these simplifications the transition probabilities read:
\begin{eqnarray}
W_{{\boldsymbol k} \uparrow \rightarrow {\boldsymbol k}^{\prime} \downarrow}^{(1)}
&=&  V^2 \gamma^2  \frac{  L^2 }{\Delta(k_{\rm F})^2}
,\label{eq-wupdown3}
\\
W_{{\boldsymbol k} \uparrow \rightarrow {\boldsymbol k}^{\prime} \uparrow}^{(1)}
&=& V^2 \frac{\left[4\beta L^2+\alpha \left( \Delta(k_{\rm F}) +\Delta \right)^2  \right]^2}{4\Delta(k_{\rm F})^2 \left( \Delta(k_{\rm F}) +\Delta \right)^2}
\label{eq-wupup3}.
\end{eqnarray}
There is no spin relaxation ($W_{{\boldsymbol k} \uparrow \rightarrow {\boldsymbol k}^{\prime} \downarrow} = 0$) in the absence of SOC, i.e. $L=0$.

Since the $k$-integrations in Eqs.~(\ref{gammas}), (\ref{gamma}) cannot be performed in the simplified four-band model, we use the reasonable assumption that $\Gamma_{\text{s}} \sim W_{k \uparrow \rightarrow k^{\prime} \downarrow}^{(1)}$ and $\Gamma \sim W_{k \uparrow \rightarrow k^{\prime} \uparrow}^{(1)}$ which will be justified by the subsequent numerical calculations.
As a result, the ratio $\Gamma_{s}/\Gamma$ reads:
\begin{eqnarray}
 \frac{\Gamma_{\text{s}}}{\Gamma} \approx \frac{W_{{\boldsymbol k} \uparrow \rightarrow {\boldsymbol k}^{\prime} \downarrow}^{(1)} }{W_{{\boldsymbol k} \uparrow \rightarrow {\boldsymbol k}^{\prime} \uparrow}^{(1)} }
= \frac{4\gamma^2 L^2 \left( \Delta(k_{\rm F}) +\Delta \right)^2}{\left[4\beta L^2+\alpha \left( \Delta(k_{\rm F}) +\Delta \right)^2  \right]^2}
.\nonumber\\
\label{eq-ratio}
\end{eqnarray}
In the limit of small SOC, i.e $L \ll \Delta$, the ratio becomes:
\begin{eqnarray}
 \frac{\Gamma_{\text{s}}}{\Gamma}  \approx \frac{\gamma^2}{\alpha^2} \frac{L^2}{\Delta^2} + {\cal O}(L^4).
\end{eqnarray}
When neglecting the constants near unity, it leads to:
\begin{eqnarray}
\Gamma_{\text{s}} \approx  \frac{L^2}{\Delta^2} \Gamma
\end{eqnarray}
reproducing the perturbation result of Elliott given in Eq.~(\ref{elliottrel1}).
In this limit, the spin-relaxation rate is much smaller than the momentum-relaxation rate.

In the opposite limit, i.e. $L \gg \Delta$, the ratio tends to a constant as a function of $L$:
\begin{eqnarray}
\frac{\Gamma_{\text{s}}}{\Gamma} \approx \frac{\gamma^2}{(\alpha+\beta)^2} + {\cal O}(1/L)
,\label{eq-RlargeSOC}
\end{eqnarray}
which means that the spin-relaxation rate approaches to the momentum-relaxation rate (apart from the constants $\alpha$, $\beta$, and $\gamma$, which are near unity).

\begin{figure}
\centering
\includegraphics[scale=.65]{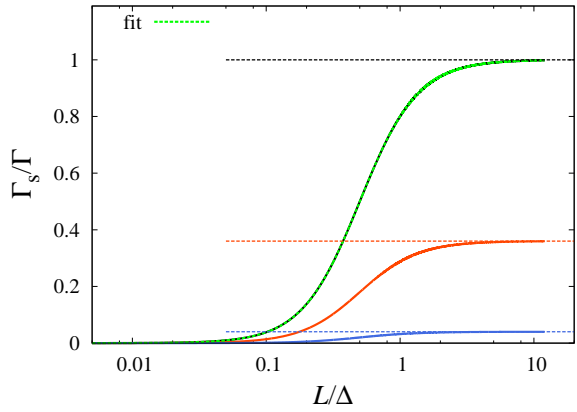}
\caption{The ratio $\Gamma_{\text{s}}/\Gamma$ as a function of the SOC matrix element $L$ with the parameter choice $\Delta=1$, $V=1$, $\alpha=0.5$, $\beta=0.5$, $\gamma=1$ ({\sl black}), $\gamma=0.6$ ({\sl red}), and $\gamma=0.2$ ({\sl blue}). {\sl Horizontal dashed lines} show the large-$L$ limits given by $\gamma^2/(\alpha+\beta)^2$. The perfect fit of the upper curve with the function in Eq.~(\ref{eq-Gs3}) is shown by {\sl dashed green line}.}
\label{fig-ratio}
\end{figure}

Fig.~\ref{fig-ratio} shows the ratio $\Gamma_{\text{s}}/\Gamma$ as a function of the SOC matrix element in the entire $L$-range with different values of the band-structure dependent constants.
The characteristic energy where the behavior of the ratio changes from $L^2$-dependence to saturation is approximately $L \approx \Delta$, i.e. where $L / \Delta \approx 1$. Fig.~\ref{fig-ratio}. also shows a fit with the following phenomenological formula, which was found to well approximate the result in the entire parameter range:
\begin{eqnarray}
\left(\frac{\Gamma_{\text{s}} }{\Gamma}\right)^{\rm approx} =  \frac{c_1 L^2  }{ \Delta(k_{\rm F})^2 + c_2 L^2}
\label{eq-Gs3}
\end{eqnarray}
with $c_{1}$ and $c_{2}$ being constants of order of unity.
%which is first-order in $\Gamma$.
We note that the simple phenomenological formula~(\ref{eq-Gs3}) is already implied by the analytic result given in Eq.~(\ref{eq-ratio}) after some manipulations (see the Supplementary Material).

We also discuss the $g$-factor shift which is given by the expectation value of the orbital momentum in the Bloch state through the Zeeman term. A straigthforward calculation (detailed in the Supplementary Material) yields for $g_{\perp}$ and $g_{\|}$, i.e. when the magnetic field is perpendicular or parallel to the $z$ axis defined by the SOC Hamiltonian (Eq. \eqref{ham-so}), respectively:

\begin{eqnarray}
g_{\perp} &=& g_0 +  \frac{L}{\sqrt{\Delta^2 + 4L^2}} ,
\label{eq-gperp}
\\
g_{\|} &=& g_0 \frac{\Delta}{\sqrt{\Delta^2 + 4L^2}}.
\label{eq-gpar}
\end{eqnarray}
The EY theory predicts an isotropic $g$-factor and the anisotropy in our results is due to the omission of the $L_{\uparrow \uparrow}$ and $L_{\downarrow \downarrow}$ terms in Eq. \eqref{ham-1}. To test the Elliott-relation we proceed with $g_{\perp}$ as:

\begin{eqnarray}
\Delta g = g_{\perp} - g_{0} = \frac{L}{\sqrt{\Delta^2 + 4L^2}},
\label{eq-g_shift}
\end{eqnarray}
which becomes $\Delta g \sim L/\Delta$ for $L \ll \Delta$ reproducing the Elliott result.

Combining expression~(\ref{eq-g_shift}) with (\ref{eq-Gs3}) gives:
\begin{eqnarray}
\Gamma_{\text{s}} \approx \Delta g^2 \Gamma,
\label{elliottrel3n}
\end{eqnarray}
which recovers the Elliott relation given in Eq.~(\ref{elliottrel3}) for arbitrary values of $L$.

\subsection*{Pseudopotential electron model of spin relaxation}

To complement the results of the model Hamiltonian calculations, we consider a nearly-free electron model where spin and momentum scattering can be calculated. The Hamiltonian of the problem is:
\begin{subequations}
\begin{eqnarray}
{\cal H} &=& {\cal H}_{\rm kin} + {\cal H}_{\rm pot} + {\cal H}_{\rm SOC};\label{ham}\\
{\cal H}_{\rm kin} &=&  \frac{p^2}{2m} = - \frac{\hbar^2}{2m} \nabla^2 ,\label{h-kin} \\
{\cal H}_{\rm pot} &=& V({\boldsymbol r}), \label{h-pot1}\\
{\cal H}_{\rm SOC} &=& \frac{\hbar^2}{4m^2c^2} (\nabla V \times {\boldsymbol k})    \cdot  {\boldsymbol \sigma}.
\label{h-so}
\end{eqnarray}
\end{subequations}
We solve it with the pseudopotential method which approximates the lattice potential $V({\boldsymbol r})$ with its first few Fourier components $V({\boldsymbol g})$ as
\begin{eqnarray}
V({\boldsymbol r})  = \sum_{\boldsymbol g} V({\boldsymbol g}) {\rm e}^{i {\boldsymbol g}\cdot {\boldsymbol r}}
.
\label{h-pot2}
\end{eqnarray}
In the calculations we take $V({\boldsymbol g}) = v({\boldsymbol g}) \cos ({\boldsymbol g} \cdot {\boldsymbol \tau})$
with ${\boldsymbol \tau}=(a/8)[1,1,1]$, which corresponds to a zincblende structure with identical atoms (example Si or Ge), i.e. with inversion symmetry retained (for further details, see Supplementary Material).

This model has some advantages and shortcomings when calculating spin relaxation. An advantage is that the SOC is obtained directly from the lattice potential thus it allows to perform a numerical analysis of spin relaxation. In addition, the pseuodpotential-based SOC automatically accounts for the crystal symmetry which is found in the SOC, too. However, the pseudopotential-based SOC is known to underestimate the experimental SOC as it considers the smooth potential of the valence electrons only and neglects the strongly oscillating potential near the atomic core\cite{yu2010}. To solve this problem, the pseudopotential-based SOC is magnified with a scaling parameter with a typical value of $\lambda_{\rm sc} \sim 10^{2}-10^{3}$. We chose $\lambda_{\rm sc} = 264$ as it account quantitatively well for the $k=0$ SOC gap in Ge\cite{yu2010}. In addition, we also introduce a \emph{tuning parameter}, $\lambda$, for the SOC interaction in order to study the SOC dependent spin relaxation. As a result, the SOC Hamiltonian reads as ${\cal H}_{\rm SOC}^{\prime} = \lambda_{\rm sc} \lambda {\cal H}_{\rm SOC}$ in Eq.~(\ref{h-so}).

\begin{figure}
\centering
\includegraphics[scale=.8]{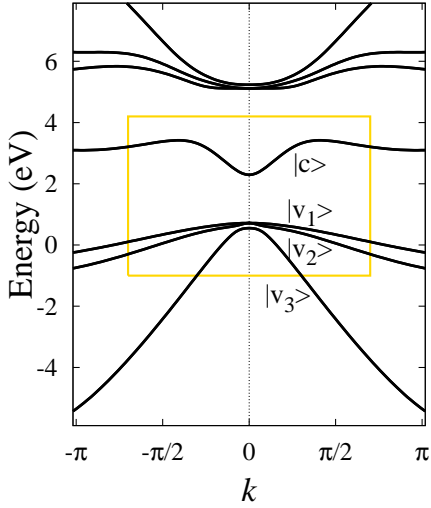}
\caption{The band structure obtained with the used pseudopotential parameters for $\lambda=1$ along the ${\boldsymbol k}=k[1/\sqrt{2},1/\sqrt{2},0]$ direction. $|v_{n} \rangle$ and $|c\rangle$ denote the valence and conduction bands (each with additional spin degeneracy $\uparrow, \downarrow$), respectively.}
\label{fig-bandstructure}
\end{figure}

We use a pseudopotential parameter set which reproduces well the  conduction and valence bands of Ge. The band structure is shown in Fig.~\ref{fig-bandstructure} for $\lambda=1$.
Considering the nearby valence and conduction bands denoted as $|v_{n}\rangle$ and $|c\rangle$ in Fig.~\ref{fig-bandstructure}, respectively, the characteristic features of metals with hybridized $s$-$p$ orbitals can be discussed. Namely, the orbital parts of the electron wave functions at ${\boldsymbol k}=0$ ($\Gamma$ point) without SOC have the symmetries $|c\rangle \sim \Gamma_{1}$ and $|v_{n}\rangle \sim \Gamma_{4}$ which mean $s$- and $p$-like orbitals, respectively. Symmetry operations acting on the spin wave functions have to be considered in the presence of SOC, which can be handled by double group irreducible representations.

At ${\boldsymbol k}=0$, the $p$-symmetric ($\Gamma_4$) valence band with spin splits into a four- and a two-fold degenerate state, separated by the SOC gap. These split states can be labeled according to the total angular momentum operator as $j=3/2 \sim \Gamma_{8}$ and $j=1/2 \sim \Gamma_{7}$, respectively. As we move away from the ${\boldsymbol k}=0$ point, the four-fold degenerate valence band splits further but a two-fold degeneracy is kept due to the presence of time-reversal symmetry. In the vicinity of the ${\boldsymbol k}=0$ point, all of the valence and conduction bands have mixed $s$-$p$ character due to the hybridization.

Using Eqs.~(\ref{wuddef1}), (\ref{wuudef1}),
we derive the non spin-flip and spin-flip transition matrix elements in the conduction band $|c\rangle$ as
\begin{eqnarray}
W_{{\boldsymbol k}\uparrow \rightarrow {\boldsymbol k}^{\prime}\uparrow}^{(c)} &=&
\delta(E_{{\boldsymbol k}} - E_{{\boldsymbol k}^{\prime}}) V^2 |_{{\boldsymbol k}}\langle c \uparrow | c \uparrow \rangle_{{\boldsymbol k}^{\prime}}|^2,
\label{eq-1cuu}\\
W_{{\boldsymbol k}\uparrow \rightarrow {\boldsymbol k}^{\prime}\downarrow}^{(c)} &=&
\delta(E_{{\boldsymbol k}} - E_{{\boldsymbol k}^{\prime}}) V^2 |_{{\boldsymbol k}}\langle c \uparrow | c \downarrow \rangle_{{\boldsymbol k}^{\prime}}|^2,
\label{eq-1cud}
\end{eqnarray}
and in the upper valence band $|v_{1}\rangle$ as
\begin{eqnarray}
W_{{\boldsymbol k}\uparrow \rightarrow {\boldsymbol k}^{\prime}\uparrow}^{(v)} &=&
\delta(E_{{\boldsymbol k}} - E_{{\boldsymbol k}^{\prime}}) V^2 |_{{\boldsymbol k}}\langle v_{1} \uparrow | v_{1} \uparrow \rangle_{{\boldsymbol k}^{\prime}}|^2,
\label{eq-1vuu}\\
W_{{\boldsymbol k}\uparrow \rightarrow {\boldsymbol k}^{\prime}\downarrow}^{(v)} &=&
\delta(E_{{\boldsymbol k}} - E_{{\boldsymbol k}^{\prime}}) V^2 |_{{\boldsymbol k}}\langle v_{1} \uparrow | v_{1} \downarrow \rangle_{{\boldsymbol k}^{\prime}}|^2
\label{eq-1vud}
\end{eqnarray}
with $V \equiv \sqrt{2\pi/\hbar} V_{{\boldsymbol k}{\boldsymbol k}^{\prime}}$ that is not specified, but the ${\boldsymbol k}$-dependence of the wave functions is obtained accurately in the pseudopotential approximation.

The momentum- and spin-relaxation rates are obtained according to Eqs.~(\ref{gammas}) and (\ref{gamma}) by averaging over the transition matrix elements numerically with respect to ${\boldsymbol k}$ and ${\boldsymbol k}^{\prime}$ on the Fermi surface with energy $E_{{\boldsymbol k}_{F}}$ as:
\begin{eqnarray}
\Gamma &=& \langle W_{{\boldsymbol k}\uparrow \rightarrow {\boldsymbol k}^{\prime}\uparrow}^{(c(v))} \rangle_{{\boldsymbol k},{\boldsymbol k}^{\prime}={\boldsymbol k}_{F}} \\
\Gamma_{\text{s}} &=& \langle W_{{\boldsymbol k}\uparrow \rightarrow {\boldsymbol k}^{\prime}\downarrow}^{(c(v))}  \rangle_{{\boldsymbol k},{\boldsymbol k}^{\prime}={\boldsymbol k}_{F}}.
\end{eqnarray}

\begin{figure}[htp]
\begin{center}
\includegraphics[scale=0.6]{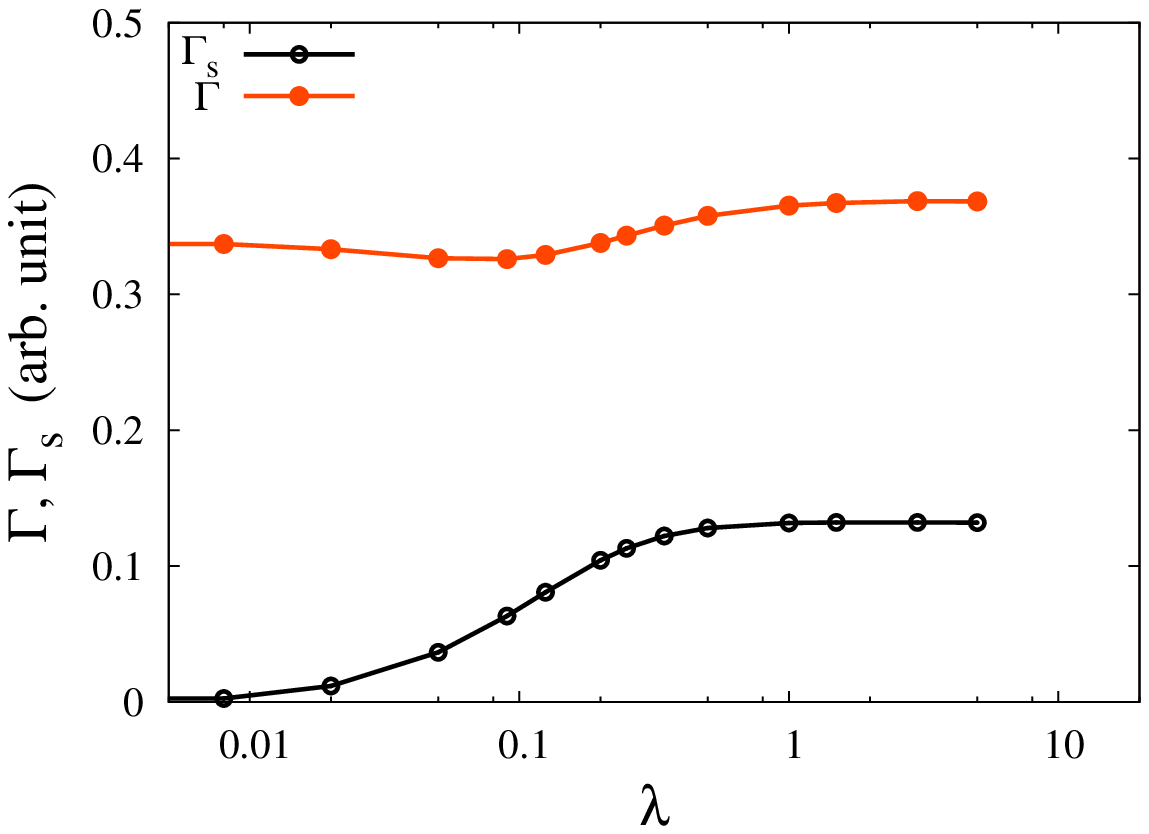}
\includegraphics[scale=0.6]{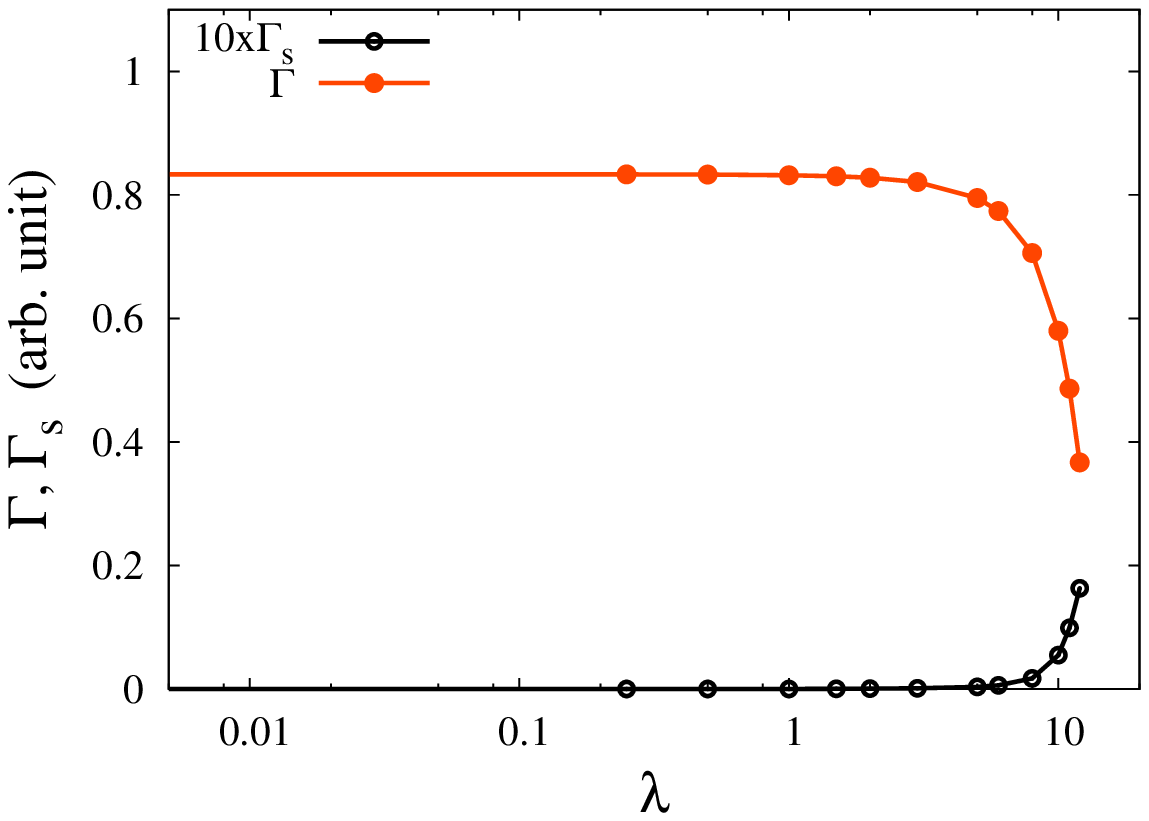}
\caption{Spin- and momentum-relaxation rates in arbitrary units, calculated in the pseudopotential approximation as a function of spin-orbit coupling $\lambda$ at $k_{\rm F}=0.4$ in the valence band ({\sl upper panel}) and in the conduction band ({\sl lower panel}).}
\label{fig-pseudoGammaGammas}
\end{center}
\end{figure}

Figure~\ref{fig-pseudoGammaGammas} shows the spin- and momentum-relaxation rates calculated in the pseudopotential approximation both for the upper valence band $|v_{1}\rangle$ and conduction band $|c\rangle$ as a function of the spin-orbit coupling, $\lambda$. We took $k_{\rm F}=0.4$ in the calculation and averaged the transition probabilities given in Eqs.~(\ref{eq-1vuu}), (\ref{eq-1vud}) over 30 points at the Fermi surface to obtain the relaxation rates, $\Gamma$ and $\Gamma_{\text{s}}$.

The seemingly different behavior of the relaxation rates in the valence and conduction bands can be consistently explained.
As it is shown in Fig.~\ref{fig-bandstructure}, the 'relevant' band gap, i.e. the characteristic energy defined as $\Delta_{\rm v} \equiv E_{v_{1}}(k_{\rm F})-E_{v_{2}}(k_{\rm F})$ for the valence state and $\Delta_{\rm c} \equiv E_{c}(k_{\rm F})-E_{v_{1}}(k_{\rm F})$ for the conduction band, is quite different for the two cases. Namely, in the case of the upper valence state, $|v_{1}\rangle$, there is another valence state $|v_{2}\rangle$ in its close vicinity, while the closest state to the conduction band $|c\rangle$ is $|v_{1}\rangle$, which is much further away, which leads to $\Delta_{\rm v} \ll \Delta_{\rm c}$.
Thus, by increasing the SOC in the calculations, the range $\lambda \gg \Delta_{\rm v}$ can be reached for the valence band and we indeed find the saturating behavior for both $\Gamma$ and $\Gamma_{\text{s}}$ in this range in well agreement with the result of the four-band model.
On the other hand, in the case of the conduction band, the relation $\lambda \ll \Delta_{\rm c}$ holds in the entire range of $\lambda$ which we study. A much larger $\lambda$ would lead to a rearranged band structure. Therefore, for the conduction band in our model, we observe the perturbative (or Elliott-Yafet) behavior $\Gamma \sim {\cal O}(1)$ and $\Gamma_{\text{s}} \sim \lambda^2$ .
%, which means that we remains at the very beginning of the curve shown in  Fig.~\ref{fig-ratio} where $L/\Delta \ll 1$.

\begin{figure}[htp]
\begin{center}
\includegraphics[scale=0.8]{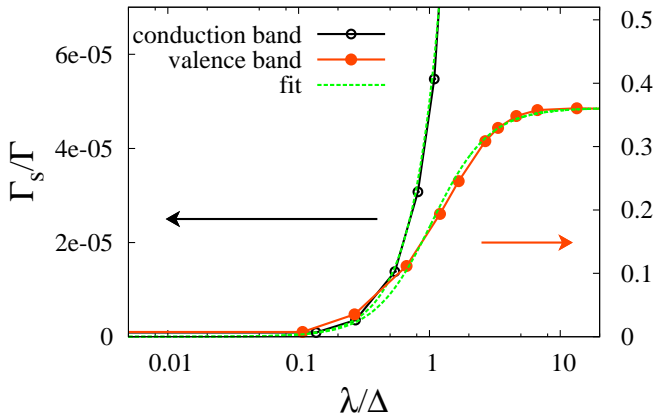}
\caption{Fit to the ratio $\Gamma_{\text{s}}/\Gamma$ calculated in the pseudopotential approximation for the valence and conduction bands as a function of SOC strength ($\lambda$) at $k_{\rm F}=0.4$ with the formula~(\ref{eq-Gs3}).}
\label{fig-pseudoratio}
\end{center}
\end{figure}

We find that the numerical data for the ratio $\Gamma_{\text{s}}/\Gamma$ follows the approximate formula obtained in the model Hamiltonian calculation for both the valence and conductions bands. Namely, the approximate formula given in Eq.~(\ref{eq-Gs3}) fits well the ratios by taking the numerical data $\Delta_{\rm v}$ and $\Delta_{\rm c}$ for the band gap $\Delta(k_{\rm F})$ for the valence and conduction band, respectively.
%Namely, the approximate formula}
%\begin{eqnarray}
%\st{\left( \frac{\Gamma_{\text{s}}}{\Gamma} \right) \approx %^{(\rm approx)}\frac{c_1 \lambda^2  }{ \Delta(k_{\rm F})^2 + c_2 \lambda^2} \label{eq-Gs4}
%\end{eqnarray}
%fits well the ratios, where we use the numerical data for the band gap in Eq.~(\ref{eq-Gs4}) as $\Delta(k_{\rm F})=\Delta_{v}$ and $\Delta(k_{\rm F})=\Delta_{c}$ for the valence and conduction bands, respectively.
Figure~\ref{fig-pseudoratio} shows the fit for the ratio $\Gamma_{\text{s}}/\Gamma$ as a function of $\lambda/\Delta$ with this model.

The good agreement between the numerically obtained data and the phenomenological formula~(\ref{eq-Gs3}) obtained in the four-band model calculation supports the previous assumption that $\Gamma_{\text{s}} \sim W_{k \uparrow \rightarrow k^{\prime} \downarrow}^{(1)}$ and $\Gamma \sim W_{k \uparrow \rightarrow k^{\prime} \uparrow}^{(1)}$. In addition, we find it compelling that the two different approaches result in $\Gamma_{\text{s}}/\Gamma$ ratios as a function of the SOC which are both well approximated by the formula in Eq. \eqref{eq-Gs3}. This as we believe, summarizes the Elliott-Yafet theory generalized for the case of strong spin-orbit coupling.

\subsection*{Comparison with experiment}
Among all the metals where experimental data on spin relaxation exists, Au is particularly suited to test the validity of the above discussion as it has the strongest SOC. In addition, gold has a single conduction electron per unit cell, i.e. its Fermi surface does not extend beyond the first Brillouin zone. It is known that description of spin relaxation is more complicated for metals with two or more conduction electrons per unit cell\cite{FabianPRL1998,FabianPRL1999} (e.g. Mg or Al), where the Fermi surface crosses the Brillouin zone boundaries giving rise to the so-called spin relaxation "hot-spots". When hot-spots are present, the Elliott relation is known to break down even though the Elliott-Yafet theory remains valid\cite{FabianPRL1998,FabianPRL1999}.
%Therein the scaling of $\Gamma_{\text{s}}/\Gamma \propto \left( L/\Delta\right)^2$ (with $L$ being the SOC matrix element) breaks down.

Monod and Beuneu found \cite{MonodBeuneuPRB1979} the SOC strength in Au as $L/\Delta \approx 0.9$. It is therefore an appropriate candidate for a case where the perturbative treatment of the SOC breaks down. Monod and J\'anossy reported in Ref. \onlinecite{MonodJanossy1977} the electron spin resonance linewidth, $\Delta B$, which is used to obtain $\Gamma_{\text{s}}$. $\Gamma$ is calculated from the resistivity, $\rho$, data in Ref. \onlinecite{Gold_resistivity} through: $\Gamma=\hbar \epsilon_0 \omega_{\text{pl}}^2 \rho$, where $\epsilon_0$ is the vacuum permittivity and $\omega_{\text{pl}}=8.55\,\text{eV}$ is the plasma frequency of gold \cite{Blaber}. For both sets of data, the residual linewidth and resistivity was subtracted as both quantities are known to obey the
Matthiessen's rule \cite{SimonPRL2008}, i.e. that the respective residual and temperature dependent scattering rates are additive.

\begin{figure}[htp]
%\begin{center}
\includegraphics[scale=.45]{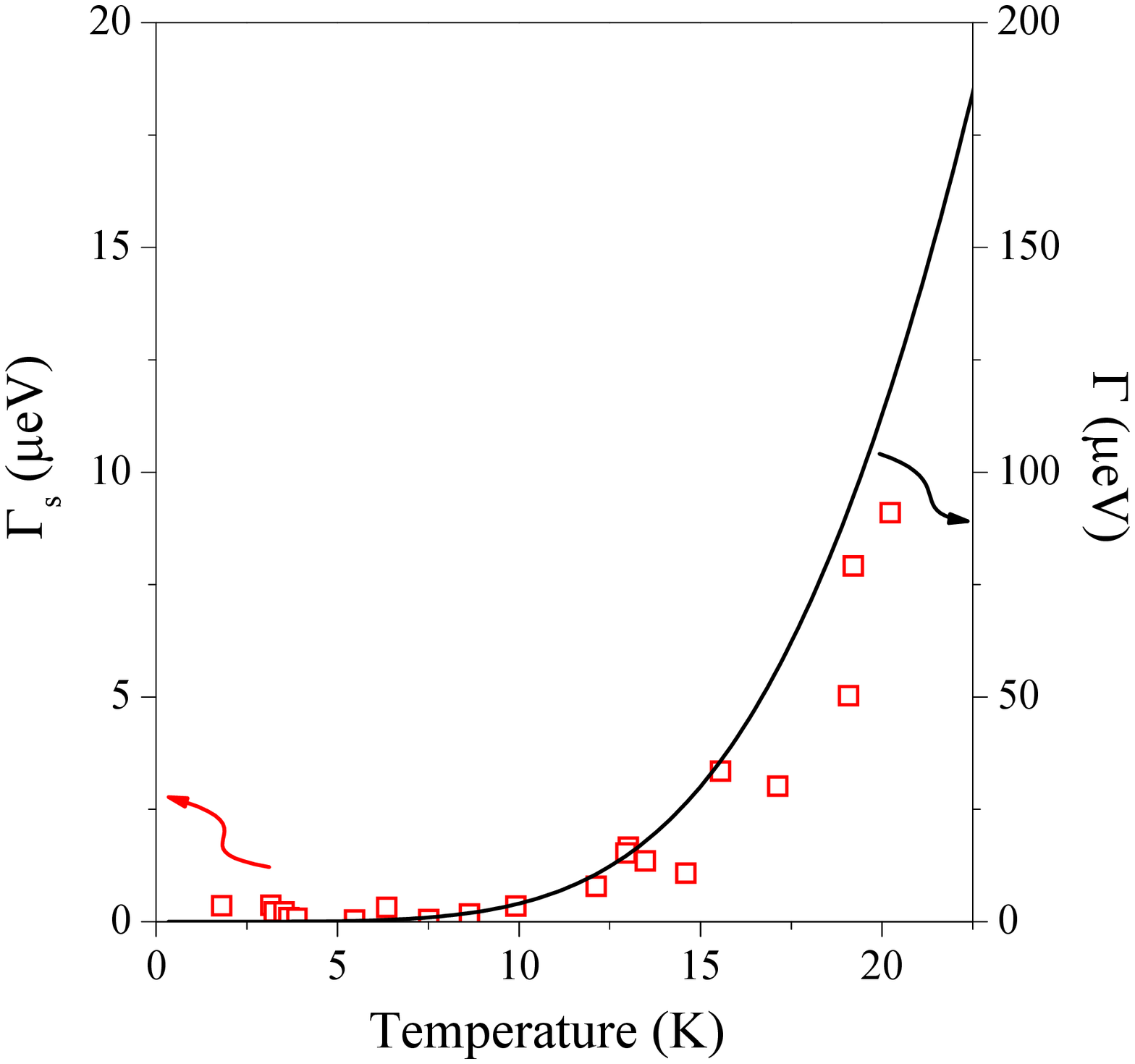}
\caption{Experimentally determined spin-lattice relaxation rate, $\Gamma_{\text{s}}$, in Au from Ref. \onlinecite{MonodJanossy1977} as compared to the momentum scattering rate, $\Gamma$. Note the factor 10 difference in the energy scales for the two quantities.}
\label{gold_relaxation}
%\end{center}
\end{figure}

The comparison, shown in Fig.~\ref{gold_relaxation}, shows that the measured spin-relaxation rate approaches the momentum-relaxation rate within an order of magnitude such as our theoretical result suggests. The difference between the predicted $\Gamma_{\text{s}}/\Gamma\approx 1$ and the observed $\Gamma_{\text{s}}/\Gamma = 0.06\pm 0.01$ could be due to the band structure dependent constants near unity in Eq. \eqref{eq-RlargeSOC}. However the observation of a nearly equal $\Gamma_{\text{s}}$ and $\Gamma$ is itself surprising as to our knowledge it has not been noted elsewhere that the mergence of spin and momentum relaxation times could be realized for a real material.

We finally test empirically the validity of the Elliott relation in Eq. \eqref{elliottrel3n}, i.e. $\Gamma_{\text{s}}/\left(\Gamma \Delta g^2\right)\approx 1$ in gold. Monod and J\'anossy found $g=2.11\pm 0.01$ which gives $\Gamma_{\text{s}}/\left(\Gamma \Delta g^2\right)\approx 5$, which is again in agreement with the theoretical prediction given the presence of the band structure dependent constant factors. This means that the Elliott relation remains a valid empirical tool to test whether the Elliott-Yafet theory applies for a given system.

\section*{Conclusions}
We generalized the Elliott-Yafet theory of spin relaxation in metals with inversion symmetry for the case of arbitrary value of the spin-orbit coupling. We applied two different approaches, exact diagonalization of a model four-band Hamiltonian (without dispersion) and numerical calculation of the spin relaxation time for a pseudopotential approach. The two methods give a qualitatively similar result, which is summarized in Eq. \eqref{eq-Gs3}. A calculation of the $g$-factor in the four-band model shows that the empirical Elliott-relation, which links the $g$-factor and spin-relaxation rate, is retained even for strong SOC. Our result predicts that spin and momentum relaxation times can have similar orders of magnitude. We show that this situation has been already observed in Au. Our result is an important step toward the unified theory of spin relaxation including the strength of the spin-orbit coupling.

\section*{Acknowledgements}
Work supported by the ERC Grant Nr. ERC-259374-Sylo and by the Hungarian Scientific Research Funds No. K106047. A. K. acknowledges the Bolyai Program of the Hungarian Academy of Sciences.

\section*{Author Contributions}
AK performed the calculations, LS presented the group theoretical considerations of the SOC, and FS discussed the comparison between the theory and experiment. All authors contributed to writing the manuscript.

\section*{Additional Information}
\textbf{Competing financial interests:} The authors declare no competing financial interests.

%\bibliographystyle{apsrev4}
%\bibliographystyle{apsrev}
%\bibliographystyle{naturemag}
%\bibliography{Tubes2011June_new}

\newpage

\appendix

\section{Group theoretical considerations on the SOC matrix elements}\label{app:group_th}

\renewcommand{\theequation}{A$\cdot$\arabic{equation}}
\setcounter{equation}{0}

\newcommand{\R}{\mathbb{R}}
\newcommand{\C}{\mathbb{C}}
\newcommand{\vect}[1]{\mathbf{#1}}
\newcommand{\conj}[1]{{#1}^{\ast}}

The Hamiltonian of the spin-orbit interaction given in Eq.~(\ref{ham-so}) in the main paper is restricted by the following three properties: i) it is self adjoint, ii) it can be written in the form of $H_1\otimes \sigma_x + H_2\otimes \sigma_y + H_3\otimes \sigma_z$ (for every $i$m $H_i$ is a kinetic operator ($2\times2$) and $\otimes$ denotes tensorial product), and iii) it is invariant to the product of inversion and time-reversal operations.

The effect of inversion ($R_i$) and time-reversal ($R_\tau$) operations on our basis are given by:

\begin{equation}
	\begin{aligned}
	R_i R_\tau \left|1 \uparrow\right> &= \left<1 \downarrow \right| , \\
	R_i R_\tau \left|1 \downarrow\right> &= -\left<1 \uparrow\right| , \\
	R_i R_\tau \left|2 \uparrow\right> &= -\left<2 \downarrow\right| , \\
	R_i R_\tau \left|2 \downarrow\right> &= \left<2 \uparrow \right| ,
	\end{aligned}
\end{equation}

\noindent where $R_i R_\tau$ denotes the product of inversion and time reversal.

Note that time reversal allows arbitrary $e^{i \varphi_{1,2}}$ factors for the two kinetic basis states, $\left|1\right>$ and $\left|2\right>$, depending how the basis were chosen. Using the notation used in main text, the second property yields $L_{\uparrow \uparrow}=-L_{\downarrow \downarrow}$. The third property implies:

\begin{equation}
	\begin{aligned}
		\left<1\uparrow\right|\hat{\cal H}_{\rm SOC}\left|2\uparrow\right>&=
		-\left<2\downarrow\right|\hat{\cal H}_{\rm SOC}\left|1\downarrow\right> , \\
		\left<1\uparrow\right|\hat{\cal H}_{\rm SOC}\left|2\downarrow\right>&=
		\left<2\uparrow\right|\hat{\cal H}_{\rm SOC}\left|1\downarrow\right>,
	\end{aligned}
\end{equation}
which in our notation for the matrix elements reads:

\begin{equation}
	\begin{aligned}
		L_{\uparrow\uparrow}&=-L_{\downarrow\downarrow}^{\ast} , \\
		L_{\downarrow\uparrow}&=L_{\uparrow\downarrow}^{\ast}.
	\end{aligned}
\end{equation}

This, together with the second restriction yields the result quoted in the main text:

\begin{equation}
	\begin{aligned}
		L_{\uparrow\uparrow}&=-L_{\downarrow\downarrow} & &\in \R, \\
		L_{\downarrow\uparrow}&=L_{\uparrow\downarrow}^{\ast} & &\in \C.\\
	\end{aligned}
\end{equation}

\section{Details of the calculations in the four-band model}\label{app:four_band}

\renewcommand{\theequation}{B$\cdot$\arabic{equation}}
\setcounter{equation}{0}

The coefficients of the mixed spin states in Eqs.~(\ref{eq-nst1}) and (\ref{eq-nst2}) of the main paper are obtained as
\begin{widetext}
\begin{eqnarray}
a_{\boldsymbol k} &=& -\frac{1}{\sqrt{2}}  \left(1 + \frac{\Delta}{\sqrt{\Delta^2 + 4 |L_{\boldsymbol k}|^2}} \right)^{1/2}
= -\frac{1}{\sqrt{2}}  \left(1 + \frac{\Delta}{\Delta({\boldsymbol k})} \right)^{1/2}
,\\
b_{\boldsymbol k} &=& \frac{1}{\sqrt{2}} \frac{L_{\boldsymbol k}^{\ast}}{|L_{\boldsymbol k}|}  \left(1 - \frac{\Delta}{\sqrt{\Delta^2 + 4 |L_{\boldsymbol k}|^2}} \right)^{1/2}
= \frac{1}{\sqrt{2}} \frac{L_{\boldsymbol k}^{\ast}}{|L_{\boldsymbol k}|} \left(1 - \frac{\Delta}{\Delta({\boldsymbol k})} \right)^{1/2}
.
\end{eqnarray}
\end{widetext}

The spin-flip and non spin-flip transition elements in the lowest band calculated from the eigenstates read:
\begin{widetext}
\begin{eqnarray}
W_{k \uparrow \rightarrow k^{\prime} \downarrow}^{(1)} &=&  \frac{2\pi}{\hbar}
\delta(E_{{\boldsymbol k}} - E_{{\boldsymbol k}^{\prime}})
\left(V_{{\boldsymbol k}{\boldsymbol k}^{\prime}} \right)^2
|_{\boldsymbol k}\langle 1 \widetilde{\uparrow} | 1 \widetilde{\downarrow} \rangle_{{\boldsymbol k}^{\prime}} |^2
=   \frac{2\pi}{\hbar} \delta(E_{{\boldsymbol k}} - E_{{\boldsymbol k}^{\prime}}) \left(V_{{\boldsymbol k}{\boldsymbol k}^{\prime}} \right)^2 \left|  a_{\boldsymbol k}^{\ast}b_{\boldsymbol k^{\prime}}^{\ast}  \,_{\boldsymbol k}\langle 1  | 2  \rangle_{\boldsymbol k^{\prime}} + b_{\boldsymbol k}^{\ast}a_{\boldsymbol k^{\prime}}^{\ast}  \,_{\boldsymbol k}\langle 2  | 1  \rangle_{\boldsymbol k^{\prime}}  \right|^2 \nonumber\\
&=&   \frac{2\pi}{\hbar} \delta(E_{{\boldsymbol k}} - E_{{\boldsymbol k}^{\prime}}) \left(V_{{\boldsymbol k}{\boldsymbol k}^{\prime}} \right)^2\frac{| \gamma_{{\boldsymbol k}{\boldsymbol k}^{\prime}} L_{\boldsymbol k^{\prime}} \left( \Delta(\boldsymbol k) +\Delta \right) + \gamma_{{\boldsymbol k}^{\prime}{\boldsymbol k}}^{\ast} L_{\boldsymbol k} \left(\Delta(\boldsymbol k^{\prime}) +\Delta \right) |^2 }{\Delta(\boldsymbol k)\Delta(\boldsymbol k^{\prime}) \left( \Delta(\boldsymbol k) +\Delta \right)\left( \Delta(\boldsymbol k^{\prime}) +\Delta \right)}
,\label{eq-wupdown1}
\\
W_{k \uparrow \rightarrow k^{\prime} \uparrow}^{(1)} &=&  \frac{2\pi}{\hbar}
\delta(E_{{\boldsymbol k}} - E_{{\boldsymbol k}^{\prime}})
\left(V_{{\boldsymbol k}{\boldsymbol k}^{\prime}} \right)^2
|_{\boldsymbol k}\langle 1 \widetilde{\uparrow} | 1 \widetilde{\uparrow} \rangle_{{\boldsymbol k}^{\prime}} |^2
=  \frac{2\pi}{\hbar}  \delta(E_{{\boldsymbol k}} - E_{{\boldsymbol k}^{\prime}}) \left(V_{{\boldsymbol k}{\boldsymbol k}^{\prime}} \right)^2 \left|  a_{\boldsymbol k}^{\ast}a_{\boldsymbol k^{\prime}}  \,_{\boldsymbol k}\langle 1  | 1  \rangle_{\boldsymbol k^{\prime}} + b_{\boldsymbol k}^{\ast}b_{\boldsymbol k^{\prime}}  \,_{\boldsymbol k}\langle 2  | 2  \rangle_{\boldsymbol k^{\prime}}  \right|^2 \nonumber\\
&=&  \frac{2\pi}{\hbar} \delta(E_{{\boldsymbol k}} - E_{{\boldsymbol k}^{\prime}}) \left(V_{{\boldsymbol k}{\boldsymbol k}^{\prime}} \right)^2 \frac{| 4\beta_{{\boldsymbol k}{\boldsymbol k}^{\prime}} L_{\boldsymbol k} L_{\boldsymbol k^{\prime}}^{\ast} +\alpha_{{\boldsymbol k}{\boldsymbol k}^{\prime}}\left( \Delta(\boldsymbol k) +\Delta \right) \left(\Delta(\boldsymbol k^{\prime}) +\Delta \right) |^2}{4\Delta(\boldsymbol k)\Delta(\boldsymbol k^{\prime}) \left( \Delta(\boldsymbol k) +\Delta \right)\left( \Delta(\boldsymbol k^{\prime})+\Delta \right)}
\label{eq-wupup1}
,
\end{eqnarray}
\end{widetext}
where we introduced the notations $_{\boldsymbol k}\langle 1  |  1  \rangle_{\boldsymbol k^{\prime}}\equiv \alpha_{{\boldsymbol k}{\boldsymbol k}^{\prime}}$, $_{\boldsymbol k}\langle 2  | 2  \rangle_{\boldsymbol k^{\prime}} \equiv \beta_{{\boldsymbol k}{\boldsymbol k}^{\prime}}$, and $_{\boldsymbol k}\langle 1  | 2  \rangle_{\boldsymbol k^{\prime}}\equiv \gamma_{{\boldsymbol k}{\boldsymbol k}^{\prime}}$ for the unknown overlap of states with different wave vectors ${\boldsymbol k}, {\boldsymbol k}^{\prime}$.

With the assumptions detailed in the main text, we obtain for the ratio of spin- and momentum-relaxation rates as:

\begin{eqnarray}
 \frac{\Gamma_{\text{s}}}{\Gamma} \approx \frac{W_{{\boldsymbol k} \uparrow \rightarrow {\boldsymbol k}^{\prime} \downarrow}^{(1)} }{W_{{\boldsymbol k} \uparrow \rightarrow {\boldsymbol k}^{\prime} \uparrow}^{(1)} }
= \frac{4\gamma^2 L^2 \left( \Delta(k_{\rm F}) +\Delta \right)^2}{\left[4\beta L^2+\alpha \left( \Delta(k_{\rm F}) +\Delta \right)^2  \right]^2}
.\nonumber\\
\label{eq-ratio_suppl}
\end{eqnarray}
This can be rewritten as:
\begin{eqnarray}
\frac{\Gamma_{\text{s}} }{\Gamma}
&=&  \frac{4 \gamma^2   L^2  }{\left[ \alpha^2 \left( \Delta(k_{\rm F}) +\Delta \right)^2 +8 \alpha \beta L^2 + \frac{(4\beta)^2 L^4}{\left( \Delta(k_{\rm F}) +\Delta \right)^2} \right]}.
\nonumber\\
\label{eq-Gs2}
\end{eqnarray}
The last term in the denominator of Eq.~(\ref{eq-Gs2}) is proportional to $L^4$ for $L \ll \Delta$ in which limit it can be neglected because the leading SOC term is $L^2$, and proportional  to $L^2$ for $L \gg \Delta$ where it can be summed up together with the original $L^2$ term. This justifies the form of the empirical fitting function for $\Gamma_{\text{s}}/\Gamma$ in the main text.

The $g$-factor is obtained from the Zeeman Hamiltonian containing both the spin and orbital term with the respective operators ($S$ and $L$):
\begin{eqnarray}
{\cal H}_{\rm Z} &=& - \mu_{\rm B} {\boldsymbol B} \cdot ({\boldsymbol L} + g_0 {\boldsymbol S})\nonumber\\
&=& - \mu_{\rm B} \left( \frac{1}{2} \left[ B_{+} (L_{+}+g_0 S_{+}) + B_{-} (L_{-}+g_0 S_{-})\right]  \right. \nonumber\\
&+& \left. B_{z} (L_{z}+g_0 S_{z})  \right)
.\label{eq-zeeman}
\end{eqnarray}
The spin-flip SOC matrix element in our four-band model connects the states $|1\sigma \rangle$ and $|2\sigma^{\prime} \rangle$, which means that only the orbital momentum operators $L_{+}$ and $L_{-}$ are to be considered in Eq.~(\ref{eq-zeeman}).

Let us change the direction of the magnetic field in the $x-z$ plane from the $z$ axis, which gives $H_{z}=H\cos \theta$, and $H_{x}=H\sin \theta$.
The expectation values of the Zeeman term between the states $|1 \widetilde{\uparrow} \rangle$ and $|1 \widetilde{\downarrow} \rangle$ given in Eqs.~(\ref{eq-nst1}), (\ref{eq-nst2}) in the main paper are calculated as
\begin{eqnarray}
\langle 1 \widetilde{\sigma} | {\cal H}_{\rm Z} | 1 \widetilde{\sigma} \rangle &=& - \langle 1 \widetilde{\sigma}^{\prime} | {\cal H}_{\rm Z} | 1 \widetilde{\sigma}^{\prime} \rangle \nonumber\\
 &=& - \mu_{\rm B} g_{0} H \cos \theta (a_{\boldsymbol k}^2 - b_{\boldsymbol k}^2),\\
\langle 1 \widetilde{\sigma} | {\cal H}_{\rm Z} | 1 \widetilde{\sigma}^{\prime} \rangle &=&
- \mu_{\rm B}  H \sin \theta \left[ g_{0}(a_{\boldsymbol k}^2 + b_{\boldsymbol k}^2) - a_{\boldsymbol k} b_{\boldsymbol k}\right].
\end{eqnarray}
Diagonalizing this $2\times 2$ problem, we obtain the modified Zeeman energies as
\begin{eqnarray}
\varepsilon_{\rm Z} = \pm \mu_{\rm B} H \sqrt{g_{\perp}^2 \sin^2 \theta + g_{\|}^2 \cos^2 \theta}
\end{eqnarray}
with
\begin{eqnarray}
g_{\perp} &=& g_0  +  \frac{L}{\sqrt{\Delta^2 + 4L^2}},
\label{eq-gperp_suppl}
\\
g_{\|} &=& g_0 \frac{\Delta}{\sqrt{\Delta^2 + 4L^2}}.
\label{eq-gpar_suppl}
\end{eqnarray}

\section{Details of the calculations in the pseudopotential electron model}\label{app:pseudo}

\renewcommand{\theequation}{C$\cdot$\arabic{equation}}
\setcounter{equation}{0}

We take the first five reciprocal vectors ${\boldsymbol g}$:
$\left[{\boldsymbol g}_{0}(1), \{ {\boldsymbol g}_{3}(8) \}, \{ {\boldsymbol g}_{4}(6) \}, \{ {\boldsymbol g}_{8}(12) \}, \{ {\boldsymbol g}_{11}(24) \}\right]$, where the degeneracy of each reciprocal vector is indicated,
 in the Fourier expansion of the lattice potential% given in Eq.~(\ref{vg})
, which leads to 102 lowest-lying states ($51\times 2$, where 2 corresponds to the spin degrees of freedom)\cite{yu2010}.

Taking the electron wave functions given in Eq.~(\ref{eq-wavef1}) of the main paper, the Schr\"odinger equation simplifies for each ${\boldsymbol k}$ to the eigenvalue problem of the Hamiltonian matrix
\begin{eqnarray}
{\cal H}_{{\boldsymbol g} \sigma, {\boldsymbol g}^{\prime}\sigma^{\prime}}({\boldsymbol k}) &=&
\left({\cal H}_{\rm kin}\right)_{{\boldsymbol g}\sigma, {\boldsymbol g}^{\prime}\sigma^{\prime}}({\boldsymbol k})
+
\left({\cal H}_{\rm pot}\right)_{{\boldsymbol g}\sigma, {\boldsymbol g}^{\prime}\sigma^{\prime}}({\boldsymbol k})
\nonumber\\
&+&
\left({\cal H}_{\rm SOC}\right)_{{\boldsymbol g}\sigma, {\boldsymbol g}^{\prime}\sigma^{\prime}}({\boldsymbol k})
 \label{secular-e1}
\end{eqnarray}
obtained from Eq.~(\ref{ham}) of the main paper, where
\begin{eqnarray}
\left({\cal H}_{\rm kin}\right)_{{\boldsymbol g}\sigma, {\boldsymbol g}^{\prime}\sigma^{\prime}}({\boldsymbol k})
&=& \frac{\hbar^2}{2m} ({\boldsymbol k} + {\boldsymbol g})^{2} \delta_{{\boldsymbol g}, {\boldsymbol g}^{\prime}} \delta_{\sigma, \sigma^{\prime}}
 \\
\left({\cal H}_{\rm pot}\right)_{{\boldsymbol g}\sigma, {\boldsymbol g}^{\prime}\sigma^{\prime}}({\boldsymbol k}) &=& V({\boldsymbol g}- {\boldsymbol g}^{\prime} )\delta_{\sigma, \sigma^{\prime}}
\\
 \left({\cal H}_{\rm SOC}\right)_{{\boldsymbol g}\sigma, {\boldsymbol g}^{\prime}\sigma^{\prime}}({\boldsymbol k}) &=&
\lambda_{\rm sc} \lambda \frac{\hbar^2}{4m^2c^2} \left[ ({\boldsymbol k}+{\boldsymbol g}) \times ({\boldsymbol k} + {\boldsymbol g}^{\prime})\right]
 \nonumber\\
& \cdot & {\boldsymbol \sigma}_{\sigma, \sigma^{\prime}}
\left[  - i  V({\boldsymbol g}- {\boldsymbol g}^{\prime} ) \right] .
 \label{secular-e2}
\end{eqnarray}

The eigenstates and eigenvalues are obtained by numerical diagonalization of the $102 \times 102$ matrix ${\cal H}_{{\boldsymbol g} \sigma, {\boldsymbol g}^{\prime}\sigma^{\prime}}({\boldsymbol k}) $.

\end{document}